\documentclass[aps, prl,superscriptaddress, twocolumn,showpacs,preprintnumbers,amsmath,amssymb]{revtex4}
\usepackage{graphicx}
\usepackage{bm}
\usepackage{amssymb, latexsym, amsmath, textcomp}

\def\bk{\boldsymbol{k}}

\def\rSU2{{\rm SU}(2)}

\newcommand{\ra}{\rangle}
\newcommand{\la}{\langle}

\begin{document} \title{Projected wavefunction study of Spin-1/2 Heisenberg
model on the Kagome lattice} 
\author{Ying Ran} 
\author{Michael Hermele}
\author{Patrick A. Lee} 
\author{Xiao-Gang Wen} 
\affiliation{Department of
Physics, Massachusetts Institute of Technology, Cambridge, Massachusetts 02139}
\date{November 13, 2006} 
\begin{abstract} We perform a Gutzwiller projected
wavefunction study for the spin-1/2 Heisenberg model on the Kagome lattice to compare
energies of several spin-liquid states. The result indicates that a U(1)-Dirac
spin-liquid state has the lowest energy. Furthermore, even without variational parameters, the energy turns out to be very close to that found by exact diagonalization. We show that such a U(1)-Dirac state
represents a quantum phase whose low energy physics is governed by four
flavors of two-component Dirac fermions coupled to a $U(1)$ gauge field.  These
results are discussed in the context of recent experiments on
ZnCu$_3$(OH)$_6$Cl$_2$.  \end{abstract}
\pacs{75.10.Jm, 75.50.Ee}

\maketitle

\begin{figure}
\includegraphics[width=0.3\textwidth]{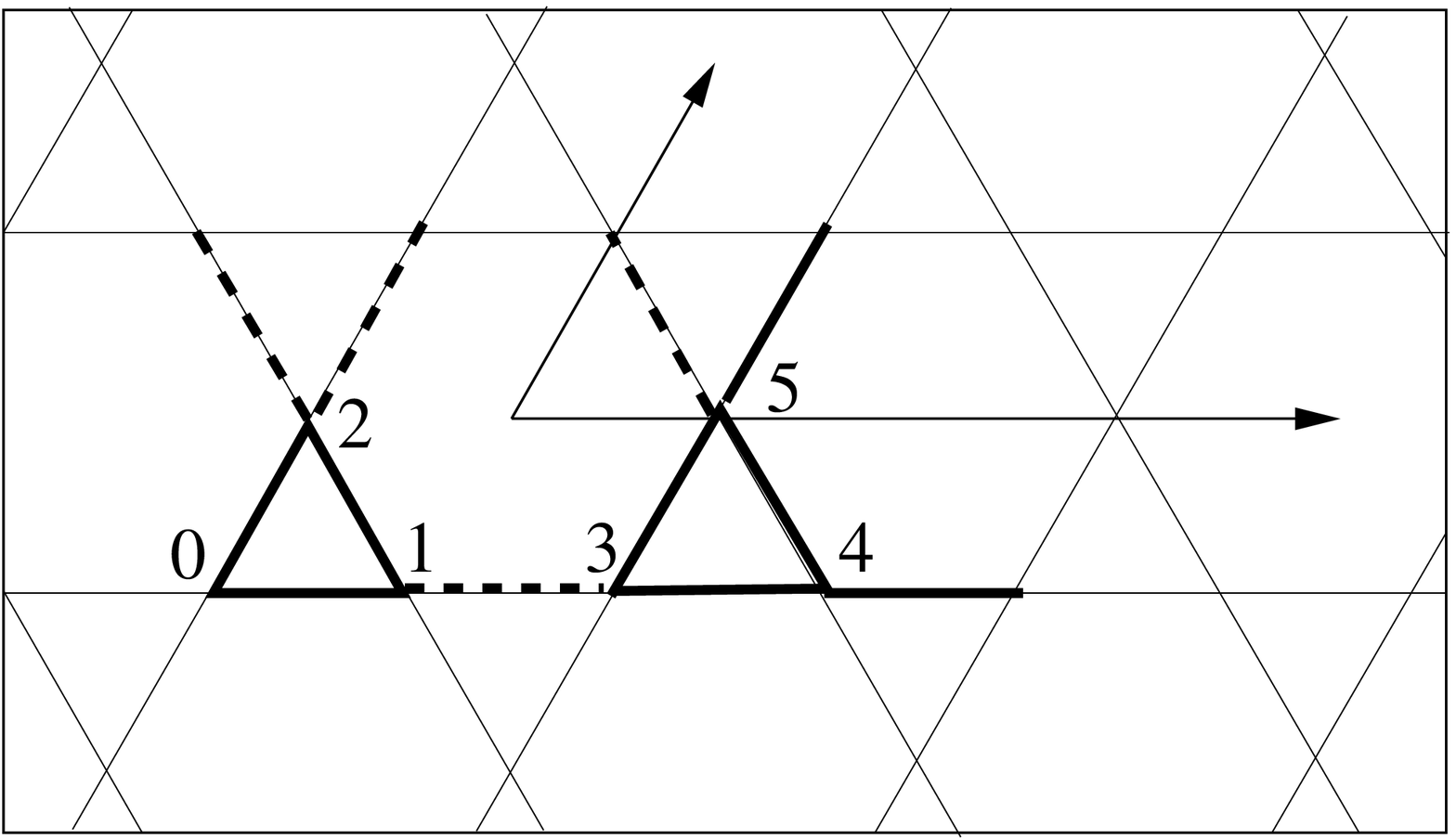}\\
\hspace{\stretch{1}}(a)\hspace{\stretch{1}}\;\\
\hspace{\stretch{1}}\includegraphics[width=0.14\textwidth]{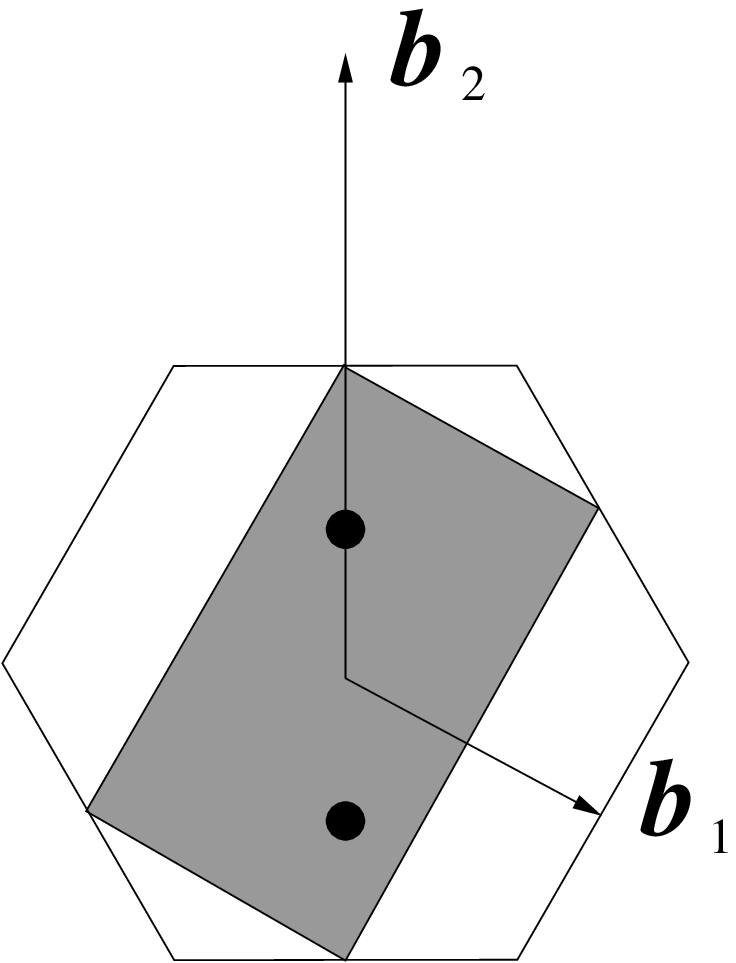}\hspace{\stretch{3}}\includegraphics[width=0.3\textwidth]{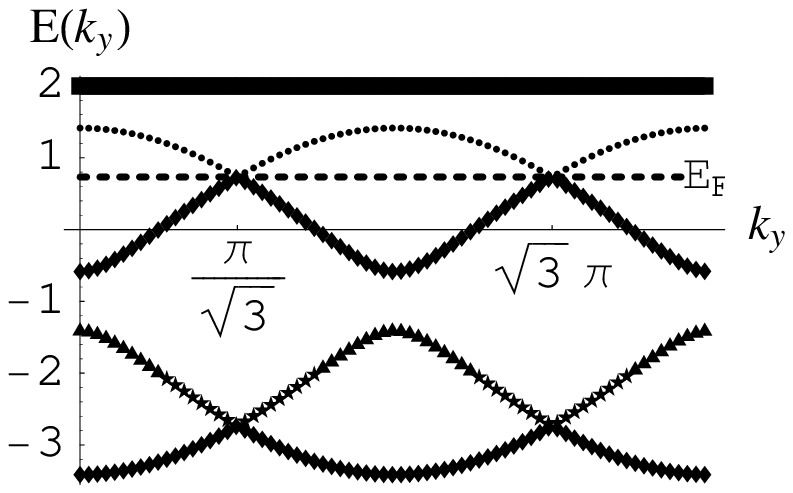}\\
\hspace{\stretch{0.4}}(b)\hspace{\stretch{1.5}}(c)\hspace{\stretch{1}}
\caption{(a): We choose the six-site unit cell for the Kagome lattice. The sites are labeled $0, \dots, 5$ as shown.  Only those bonds depicted as bold solid lines and bold dashed lines are contained within the unit cell. The bold solid bonds have positive hopping, and the bold dashed bonds have negative hopping. (b): The Brillouin zone for the doubled unit cell (gray area), with reciprocal lattice basis vectors $\boldsymbol{b}_i$ shown.  The outer hexagon is the Brillouin zone for the 3-site unit cell of a single up-pointing triangle.  The positions of the Dirac nodes are denoted by the black circles. (c): Plot of the band structure of the $U(1)$-Dirac state on the line from $\bk = 0$ to $\bk = \boldsymbol{b}_2$ with energy in units of $\chi J$(see text). The flat band is doubly-degenerate; all others are nondegenerate.  The Fermi level corresponding to one spinon per site is indicated by the dashed line.}
\label{kdbuc}
\end{figure}

Recent experimental studies of a spin-1/2 Kagome system
ZnCu$_3$(OH)$_6$Cl$_2$\cite{helton-2006,ofer-2006,mendels-2006} show that the
system is in a non-magnetic ground state. The Kagome lattice can be viewed as
corner-sharing triangles in two-dimension(Fig.\ref{kdbuc}(a)). The compound
shows no magnetic order down to very low temperature (50mK) compared with the
Curie-Weiss temperature ($>$200K). The spin susceptibility rises with
decreasing temperature, but saturates to a finite value below 0.3 K. The
specific heat is consistent with a linear T behavior below 0.5 K. There is no
sign of a spin gap in dynamical neutron scattering. These observations led us
to re-examine the issue of the ground state of the spin-1/2 Kagome lattice.

Based on Monte Carlo studies of Gutzwiller projected wavefunctions, we propose
the ground state to be a U(1)-Dirac spin-liquid state which has relativistic
Dirac spinons. The low energy effective theory is a U(1) gauge field coupled to
four flavors of two-component Dirac fermions in 2+1 dimension. This state was
studied earlier in the mean-field approximation\cite{PhysRevB.63.014413}.
However, that study focused on an instability toward a
 Valence Bond Solid (VBS) state which
breaks translation symmetry\cite{PhysRevB.63.014413}; it was not appreciated that the
U(1)-Dirac state can be a stable phase.  Using the Projective Symmetry
Group\cite{PhysRevB.65.165113,PhysRevB.70.214437,ran-2006} (PSG) technique, we
reconsider the stability of the U(1)-Dirac state and find it can be stable. Our
numerical calculations confirm that neighbor states like the VBS states and
chiral spin-liquid state all have higher energies.

One way to construct spin-liquid states is to introduce fermionic spinon
operators \cite{2986272,3246071} $f_{\uparrow}$ and $f_{\downarrow}$ to
represent the bosonic spin operator: $\vec{\mathbf{S}}_i
=\frac{1}{2}f_{i\alpha}^{\dag} \vec{\sigma}_{\alpha\beta}f^{\vphantom\dagger}_{i\beta}$.  This
representation enlarges the Hilbert space, and a local constraint is needed to
go back to the physical Hilbert space:
$f_{\uparrow}^{\dag}f^{\vphantom\dagger}_{\uparrow}+f_{\downarrow}^{\dag}f^{\vphantom\dagger}_{\downarrow}=1$.  
For the nearest neighbor Heisenberg model
\begin{align} H=J\sum_{\la ij\ra}
\vec{\mathbf{S}}_i\cdot\vec{\mathbf{S}}_j,
\label{HH} 
\end{align} 
we can substitute the spin operator by the spinon operators, so that the spin interaction is represented as a four-fermion
interaction. The four-fermion interaction can be decomposed via a
Hubbard-Stratonovich transformation by introducing the complex field $\chi_{ij}$
living on the links.  The path integral of the spin model is then $Z=\int d\chi
d\lambda d f df^{\dag}e^{-S}$, where the action is 
\begin{align}
S=&\int d\tau \Big[ \sum_{i}f_{i\alpha}^{\dag}\partial_{\tau}f^{\vphantom\dagger}_{i\alpha}+i\lambda_i(f_{i\alpha}^{\dag}f^{\vphantom\dagger}_{i\alpha}-1)\notag\\
&\sum_{ij}2J\left|\chi_{ij}\right|^2+J(\chi_{ij}f_{j\alpha}^{\dag}f^{\vphantom\dagger}_{i\alpha}+h.c.)\Big]
\end{align}
Here $\lambda$ is the Lagrangian multiplier to ensure the local constraint,
and it can be viewed as the time component of a compact U(1) gauge field,
whereas the phase of $\chi_{ij}$ can be viewed as the space components of the
same gauge field. Only when the full gauge field fluctuations are included can
one go back to the physical Hilbert space.

With this fermionic representation, one can do a mean-field study of the
spin-liquid states by taking $\chi_{ij}$'s as mean-field parameters. For the
Kagome lattice, the mean-field states are characterized by the fluxes through
the triangles and the hexagons. Controlled mean-field studies were done by
generalizing the $SU(2)$ spin model to $SU(N)$ spin model via introducing $N/2$
flavors of fermions\cite{3950277,PhysRevB.63.014413}, and several candidate
states were found: 
\begin{enumerate} 
\item\label{SL1} Valence Bond Solid (VBS)
states which breaks translation symmetry.  
\item\label{SL2} a spin liquid state
(SL-$\left[\frac{\pi}{2},0\right]$) with a flux $+\pi/2$ through each triangle
on Kagome lattice and zero-flux through the hexagons. This is a chiral spin
liquid which breaks time-reversal symmetry.  
\item\label{SL3} a spin liquid
state (SL-$\left[\pm\frac{\pi}{2},0\right]$) with staggered $\pi/2$-flux
through the triangles ($+\frac{\pi}{2}$ through up triangles and
$-\frac{\pi}{2}$ through down triangles) and zero-flux through the hexagons.
\item\label{SL4} a spin liquid state (SL-$\left[\frac{\pi}{2},\pi\right]$) with
$+\pi/2$-flux through the triangles and $\pi$-flux through the hexagons.
\item\label{SL5} a uniform RVB spin liquid state (SL-$\left[0,0\right]$) with
zero-flux through both triangles and hexagons. This state has a spinon Fermi
surface.  
\item\label{SL6} a U(1)-Dirac spin liquid state
(SL-$\left[0,\pi\right]$) with zero-flux through the triangles and $\pi$-flux
through the hexagons. This state has four flavors of two-component Dirac
fermions.  
\end{enumerate}

Marston and Zeng\cite{3950277} found that among the spin liquid states
(\ref{SL2})-(\ref{SL5}), the chiral spin liquid
SL-$\left[\frac{\pi}{2},0\right]$ has the lowerest mean-field energy. But
numerical calculations\cite{98074300284} do not support a large
chirality-chirality correlation, and Hastings\cite{PhysRevB.63.014413} found
SL-$\left[0,\pi\right]$ to be the state with the lowest mean-field energy among
the non-chiral spin liquid states. However its mean-field energy is still
higher than that of (\ref{SL2}). All the above mean-field arguments are based
on the $\frac{1}{N}$ expansion treatment of gauge fluctuation, which may fail
when $N=2$ in the physical case. To clarify which candidate is the physically
low energy spin liquid state, we do a Monte Carlo study on the trial projected
wavefunctions\cite{3381199}.

As we mentioned, fermionic representation enlarges the Hilbert space. One way
to treat the unphysical states is to include gauge fluctuations. Another direct
way to remove the unphysical states is to do a projection by hand. Suppose that
we have a mean-field ground state wavefunction $\vert \Psi_{mean}(\chi_{ij})
\ra$ with mean-field parameters $\chi_{ij}$'s, we can project out all the
unphysical states, then the resulting
wavefunction $\vert \Psi_{prj} \ra$ would be a physical state: $
\vert \Psi_{prj}(\chi_{ij}) \ra=P_D \vert \Psi_{mean}(\chi_{ij})
\ra$, where $P_D=\prod_i(1-n_{i\uparrow}n_{i\downarrow})$
is the projection operator ensuring one fermion per site. It turns out that the
calculation of energy $\la\Psi_{prj}\vert H\vert\Psi_{prj}\ra$ can be
implemented by a Monte Carlo approach with power law complexity, which means
that one can do a fairly large lattice.\cite{3381199}

We note that states related by a global transformation $\chi_{ij}\rightarrow
-\chi_{ij}^{*}$ represent the same spin wavefunction after projection. This is
because projection of holes $h_{\uparrow}=f_{\downarrow}^{\dag}$ gives the same
state as projection of fermions and the hopping parameter is transformed
accordingly. This is a special case of the $SU(2)$ gauge
symmetry\cite{9015906}. In particular, if $\chi_{ij}$ is real, the projected
state and its energy are independent of a global sign change.

For spin-1/2 nearest neighbor anti-ferromagnetic Heisenberg model
Eq.(\ref{HH}), we did the Monte Carlo calculation for energies of projected
spin liquid states on lattices with 8x8 and 12x12 unit cells (each unit cell
has 3 sites). We choose mixed boundary conditions; i.e., periodic along one Bravais lattice vector, and anti-periodic along the other Bravais lattice vector. The results are summarized
in Table \ref{result}.

\begin{table} 
\begin{tabular}{|l|l|l|} 
\hline 
Spin liquid&8x8x3 lattice&12x12x3 lattice\\ 
\hline 
SL-$\left[\frac{\pi}{2},0\right]$&-0.4010(1)&-0.4010(1)\\
\hline 
SL-$\left[\pm\frac{\pi}{2},0\right]$&-0.3907(1)&-0.3910(1)\\ 
\hline
SL-$\left[\frac{\pi}{2},\pi\right]$&-0.3814(1)&-0.3822(1)\\ 
\hline
SL-$\left[0,0\right]$&-0.4115(1)&-0.4121(1)\\ 
\hline
SL-$\left[0,\pi\right]$&-0.42866(2)&-0.42863(2)\\ 
\hline 
\end{tabular}
\caption{For all candidate projected spin-liquids, we list the energy per site
in unit of $J$. The U(1)-Dirac state SL-$\left[0,\pi\right]$ is the lowest
energy state, and its energy is even lower than some numerical estimates of the
ground state energy(see Table \ref{others}).} \label{result} 
\end{table}

We found that the U(1)-Dirac state (this state is the \emph{projected} spin
liquid of the mean-field state (\ref{SL6}) proposed by
Hastings\cite{PhysRevB.63.014413}) has the lowest energy, which is $-0.429J$
per site. Note that these results change the order of mean-field energies of
the spin liquids (\ref{SL2})-(\ref{SL6}), where
the chiral spin liquid (\ref{SL2}) was found to be of the lowest energy. In
Table \ref{others} we list the estimates of the ground state energy by various
methods. It is striking that even though the projected U(1)-Dirac state has
\emph{no variational parameter}, it has an energy which is even lower than some
numerical estimates of ground state energy. Furthermore its energy is very close to the
exact diagonalization result when extrapolated to large sample size. Thus we
propose it to be the ground state of the spin-1/2 nearest neighbor Heisenberg
model on the Kagome lattice.

\begin{table} \begin{tabular}{|l|l|} \hline Method&energy per site\\ \hline
Exact Diagonalization\cite{98074300284}&-0.43\\ \hline Coupled Cluster
Method\cite{PhysRevB.63.220402}&-0.4252\\ \hline Spin-wave Variational
method\cite{PhysRevB.69.224414}&-0.419\\ \hline \end{tabular} \caption{We list
the previous estimates for ground state energy in unit of $J$.} \label{others}
\end{table}

Hastings\cite{PhysRevB.63.014413} proposed a neighboring VBS ordered state 
as an instability of the U(1)-Dirac state.
This state can be obtained by giving the fermions non-chiral
masses. In particular, he proposed a VBS state with with a $2 \times 2$ expansion of the
unit cell. The 12 hopping parameters on the boundary of the star of David (six
triangles surounding the hexagon) have amplitude $\chi_1$, while all other
hoppings have amplitude $\chi_2$. Our numerical calculations show that this VBS
ordered state has higher energy (see Table \ref{hastings_instable_numerical}), so the U(1)-Dirac state
is stable against VBS ordering. 
 Another neighbor state
of the U(1)-Dirac spin liquid discussed by Hastings\cite{PhysRevB.63.014413} is obtained
by giving the fermions chiral masses.  The resulting state is a chiral spin liquid with broken time-reversal symmetry, and has $\theta$-flux through the
triangles and $(\pi-2\theta)$-flux through the hexagons (if $\theta=0$ the state
goes back to the U(1)-Dirac state). In Table \ref{hastings_instable_numerical}
we also show that non-zero $\theta$ increases the energy.

To determine whether the U(1)-Dirac state is a stable phase, 
we start with its effective theory 
\begin{align}
\label{effS}
S=&\int dx^3 \big[
\frac{1}{g^2}(\varepsilon_{\lambda\mu\nu}\partial_{\mu}a_{\nu})^2+\sum_{\sigma}\bar{\psi}_{+\sigma}\left(
\partial_{\mu}-ia_\mu\right)\tau_{\mu}\psi_{+\sigma}\notag\\
&+\sum_{\sigma}\bar{\psi}_{-\sigma}\left(
\partial_{\mu}-ia_\mu\right)\tau_{\mu}\psi_{-\sigma}\big]
+ \cdots ,
\end{align} 
where the first term comes from integrating out some higher energy fermions, and $\cdots$ represents other terms that are generated by interaction.
The massless Dirac fermions in the effective theory come from the gapless nodal
spinons in the mean-field theory.  The two-component Dirac spinor fields 
are denoted by $\psi_{\pm \sigma}$, where $\pm$ label the two inequivalent
nodes and $\sigma$ the up/down spins.  Also, $\bar{\psi}_{\pm \sigma} = \psi^\dagger_{\pm \sigma} \tau^3$, and the $\tau_\mu$ are Pauli matrices. The massless fermions lead to an
algebraic spin liquid \cite{PhysRevB.65.165113,PhysRevB.70.214437}.
The stability of the U(1)-Dirac state can now be determined by
examining the $\cdots$ terms: If $\cdots$ terms contain no relevant
perturbations, then the U(1)-Dirac state can be stable.

The potential relevant terms are 16 possible gauge-invariant, spin-singlet
bilinears of $\psi_{\pm \sigma}$.  To see if those bilinears are generated by
interaction or not, we need to study how lattice symmetries are realized in the
effective theory (\ref{effS}).  Because spinons are not gauge invariant, lattice symmetry is realized in the effective theory as a projective symmetry, described by a PSG.   This means that the realization of lattice symmetry includes nontrivial gauge transformations.  We find that 15 of 16
bilinears are forbidden by translation symmetry and time-reversal alone.
The remaining bilinear, which is allowed by symmetry, is $\sum_{\pm, \sigma}
\psi^\dagger_{\pm \sigma} \psi_{\pm \sigma}$.  This term shifts the spinon
Fermi level to make the ground state to have exactly one spinon per site. In
this case, the lower three of six spinon bands are filled and the spinon Fermi level
is exactly at the gapless nodal points.
This analysis tells us that the U(1)-Dirac state is stable in mean-field theory
(and also in a large-$N$ treatment). Because 
not all scaling exponents are known
in such an algebraic spin liquid,
perturbations other than fermion bilinears could in principle lead to an
instability.  However, so far, the variational wavefunction analysis suggests
that this is not the case and that the U(1)-Dirac state is stable.

\begin{table} \begin{tabular}{|l|l|l|} \hline State &8x8x3 lattice&12x12x3
lattice\\ \hline U(1)-Dirac spin liquid&-0.42866(2)&-0.42863(2)\\ \hline VBS
state($\vert \chi_1/\chi_2\vert=1.05$)&-0.42848(2)&-0.42844(2)\\ \hline VBS
state($\vert \chi_1/\chi_2\vert=0.95$)&-0.42846(2)&-0.42846(2)\\ \hline Chiral
spin liquid($\theta=0.05$)&-0.42857(2)&-0.42853(2)\\ \hline \end{tabular}
\caption{We list the energy per site in unit of $J$ for possible instabilities
of the U(1)-Dirac spin liquid. The VBS state and the chiral spin liquid are those discussed in
Ref\cite{PhysRevB.63.014413}(see text). One can see that both VBS order and chiral spin liquid increase the
energy.} \label{hastings_instable_numerical} \end{table}

Now we study the U(1)-Dirac spin liquid on the mean-field level. The U(1)-Dirac
mean-field state is defined as the ground state of the following tight-binding
spinon Hamiltonian: $H_{mean}=J\sum_{\la
ij\ra}\chi_{ij}f_{j\alpha}^{\dag}f^{\vphantom\dagger}_{i\alpha}+h.c.$. All $\chi_{ij}$'s have the
same magnitude and they produce zero-flux through the triangles and $\pi$-flux
through the hexagons.

Although the U(1)-Dirac state does not break translation symmetry (because the
translated state differs from the original state only by a gauge
transformation), the unit cell has to be doubled to work out the mean-field
spinon band structure. One can fix a gauge in which all hoppings are real as
shown in Fig.~\ref{kdbuc}(a). In this gauge the Dirac nodes are found to be at
$\boldsymbol{k}=(0,\pm\frac{\pi}{\sqrt{3}a})$ as shown in Fig.~\ref{kdbuc}(b)(c), where $a$ is
the Kagome unit cell spacing, i.e., twice the nearest neighbor distance. These
are isotropic Dirac nodes; i.e., the Fermi velocity is the same in all
directions. More nodes can be obtained by covering the momentum space by
repeating the Brillouin zones. One can easily see that the Dirac nodes form a
triangular lattice in momentum space with lattice spacing $\frac{2\pi}{\sqrt{3}a}$. Note that the positions of the Dirac nodes are gauge dependent, but the
momentum vectors connecting any two Dirac nodes are gauge invariant. Because
the spinon excitations are gapless at the nodal points, we expect the spin-1
excitations of the U(1)-Dirac spin liquid are also gapless at zero momentum and those momenta
connecting two Dirac nodes.

For the ZnCu$_3$(OH)$_6$Cl$_2$ compound, the Heisenberg coupling was estimated
to be $J\approx300$K\cite{helton-2006}, and one can calculate the Fermi
velocity at mean-field level. We find that the Fermi velocity of the U(1)-Dirac
ansatz is $v_F=\frac{a \chi J}{\sqrt{2}\hbar}$, where $\chi$ is the magnitude
of the self-consistent mean-field parameter. Hastings\cite{PhysRevB.63.014413}
found that $\chi=0.221$. $\chi$ describes the renormalization of the spinon
bandwidth and is not expected to be given quantitatively by the mean-field
theory. Hence in the formulae below we retain $\chi$ as a parameter. We find
$v_F=\frac{a\chi J}{\sqrt{2}\hbar}=19\chi\cdot 10^3 \mbox{ m}/\mbox{s}$.

We can also calculate the specific heat of the U(1)-Dirac state at mean-field
level. At low temperature ($k_B T\ll\chi J$), one expects a $C\propto T^2$ law
because of the Dirac nodes. The coefficient is related to $v_F$: 
\begin{align} 
\frac{C}{T^2}=\frac{72\zeta(3)\pi k_B^3 A}{(2\pi\hbar
v_F)^2}=1.1 \chi^{-2}\cdot 10^{-3} \mbox{ Joule}/\mbox{mol K}^3,
\label{sh}
\end{align} 
where $A$ is the area of the 2-D system. (Note that for ZnCu$_3$(OH)$_6$Cl$_2$
compound, the unit cell spacing $a=6.83$\AA, so $A=2.4 \cdot 10^5$ m$^2$/mol, where mole refers to one formula unit. We also used the fact that there are four two-component Dirac
fermions.)

In a magnetic field, the spinons will form a Fermi pocket whose radius is
proportional to magnetic field strength. Therefore at low temperature
$k_BT\ll\mu_B B$, the specific heat is linear in $T$: 
\begin{align*}
\frac{C}{T}=\frac{8\pi^3k_B^2 A\mu_B B}{3(2\pi\hbar v_F)^2}=0.23\chi^{-2}
B\cdot 10^{-3} \mbox{ Joule/mol K}^2, 
\end{align*} 
where magnetic field $B$ is
in unit of Tesla. We also find in the temperature range $\mu_B B\ll k_BT\ll\chi
J$,
\begin{align}
C&= \frac{24\pi A k_B^3 T^2}{(2\pi\hbar v_F)^2}\Big[3
\zeta(3)+\frac{2\ln2}{3}\Big(\frac{\mu_B B}{k_B T}\Big)^2
+ O(B^4)\Big].\nonumber 
\end{align}
Keeping the lowest order correction, the specific heat has a
\emph{temperature independent} increase proportional to $B^2$.  
\begin{align}
\Delta C&=\frac{16\pi\ln 2\; k_B A}{(2\pi\hbar
v_F)^2}\left(\mu_B B\right)^2
\notag\\ &
=6.3\chi^{-2}B^2\cdot
10^{-5}\;\mbox{Joule/mol K}.\label{cshift} 
\end{align}
This is in contrast to the specific heat shift of a local moment, which
decreases with $T$ as $B^2/T^2$. Eq.(\ref{cshift}) provides a way to seperate
the Dirac fermion contribution from that of impurities and phonons.

The gauge field also gives a $T^2$ contribution to the specific heat.  However, in a large-$N$ treatment this will be down by a factor of $1/N$ compared to the fermion contribution.
Furthermore, the self energy correction due to gauge fluctuations does not lead
to singular corrections to the Fermi velocity\cite{PhysRevLett.79.2109}, so the
$T^2$ dependence of $C$ is a robust prediction.

We notice that experiment observed that the specific heat of Kagome compound
ZnCu$_3$(OH)$_6$Cl$_2$  behaves as $C\propto T^{2/3}$ in zero magnetic field
over the temperature window 106 mK $<T<$ 600 mK\cite{helton-2006}, which is
enhanced from $C\propto T^2$ law. This enhancement is suppressed by a modest
magnetic field\cite{helton-2006}. Furthermore, over a large temperature range
(10K to 100K), the spin susceptibility is consistent with Curie's law with 6\%
impurity local moment. We propose that these impurity spins (possibly due to Cu
located on the Zn sites) may be coupled to the spinons to form a Kondo type
ground state with a Kondo temperature $\lesssim 1$K, thus accounting for the
large $C/T$ and the saturation of the spin susceptibility below 0.3K. The Kondo
physics of impurities coupled to Dirac spinons is in itself a novel problem
worthy of a seperate study. Meanwhile it appears to dominate the low
temperature properties and obscure the true excitations of the Kagome system.
We propose that a better place to look for the Dirac spectrum may be at higher
temperature (above 10K) and as a function of magnetic field, where the impurity
contributions may be suppressed and the unique signature of Eq.(\ref{sh}) and
Eq.(\ref{cshift}) may be tested. On the other hand, we caution that from Fig.\ref{kdbuc}(c), the spinon spectrum deviates from linearity already at a relatively low energy scale($\sim0.5\chi J$). Our theory also predicts a linear $T$ spin
susceptibility of $k_BT\ll \chi J$. Knight shift measured by Cu NMR is the
method of choice to seperate this from the impurity contribution.

Finally we remark on a possible comparison with exact diagonalization studies
which found a small spin gap of $\sim \frac{J}{20}$ and a large number of low
energy singlets\cite{98074300284}. It is not clear whether these results can be
reconciled with a $U(1)$-Dirac spin liquid. Here we simply remark that in a
finite system the Dirac nodes can easily produce a small triplet gap and that
the gauge fluctuations may be responsible for low energy singlet excitations.

We thank J. Helton and Y. S. Lee for helpful discussions. 
This research is supported by NSF grant No.  DMR-0433632 and DMR-0517222.

\bibliographystyle{apsrev}
\bibliography{/home/ranying/downloads/reference/simplifiedying}

\end{document}